\documentclass[floatfix]{jpconf}
\usepackage{graphicx}
\usepackage{amsmath}
\usepackage{amsfonts}
\usepackage{amssymb}
\usepackage{color}
\usepackage{verbatim}
\makeatletter

\newcommand{\Rmnum}[1]{\expandafter\@slowromancap\romannumeral  #1@}

\newcommand{\tcite}[1]{Ref.~\cite{#1}}

\allowdisplaybreaks

\makeatother
\begin{document}
\title{\large Nonequilibrium Green's functions and atom-surface dynamics: 
Simple views from a simple model system.
}
\date{\today}
\author{E. Bostr\"om$^1$, M. Hopjan$^1$, A. Kartsev $^2$, C. Verdozzi$^1$, and C.-O. Almbladh$^1$}
\address{$^1$ Mathematical Physics and ETSF, Lund University, Box 118, S-22100 Lund, Sweden}
\address{$^2$ National Research Tomsk State University, Lenina pr. 36, 634050 Tomsk, Russia}
\ead{emil.bostrom@teorfys.lu.se, miroslav.hopjan@teorfys.lu.se}
\begin{abstract}
We employ Non-equilibrium Green's functions (NEGF) to describe the
real-time dynamics of an adsorbate-surface model system
exposed to ultrafast laser pulses.
For a finite number of electronic orbitals, the system 
is solved exactly and within different levels of approximation. Specifically
i) the full exact quantum mechanical solution for electron and nuclear
degrees of freedom is used to benchmark ii) the Ehrenfest approximation (EA) for the nuclei, 
with the electron dynamics still treated exactly. Then, using
the EA, electronic correlations are treated with  NEGF within
iii) 2nd Born and with
iv) a recently introduced hybrid scheme, which mixes 2nd Born self-energies
with non-perturbative, local exchange-correlation potentials of Density Functional Theory (DFT). 
Finally, the effect of a semi-infinite substrate
is considered: we observe that a macroscopic number of de-excitation channels 
can hinder desorption.
While very preliminary in character and based on
a simple and rather specific model system, our results clearly illustrate the 
large potential of NEGF to investigate atomic desorption, and more generally, 
the non equilibrium dynamics of material surfaces subject to ultrafast laser fields.
\end{abstract}
\section{Introduction}
\begin{figure}[b]
       \centering
        \includegraphics[width=80mm]{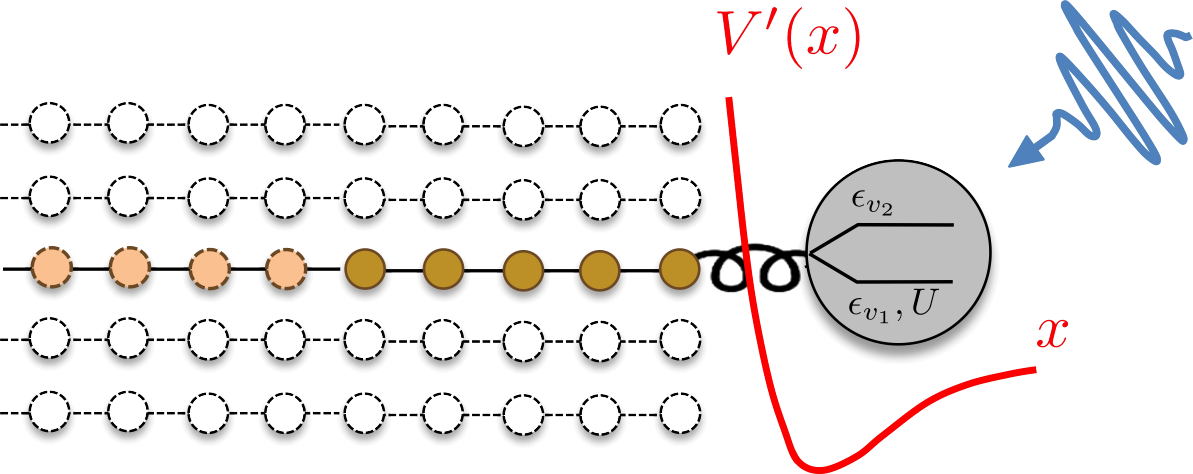}
        \label{fig:model}
\caption{The model system for adsorbate-dynamics investigated in this work, for a finite (5-site)  and infinite 1D substrate (a bulk substrate is also shown). The adsorbate is coupled to the substrate via an hopping term $V'(x)$ which depends on the adsorbate-substrate distance $x$
(a schematic effective adsorbate potential is also shown). The interaction $U$ among electrons 
is present only in the lower valence level of the adsorbate (see Eq.~(\ref{Ham0}) and afterwards for details).
}
\label{fig:system}
\end{figure}

Since the advent of femtosecond (fs) spectroscopy, the experimental characterisation of ultrafast chemical reactions has grown spectacularly \cite{Zewail}. At the same time, with the development of ever more sophisticated techniques to probe and control atomic and molecular processes far away from equilibrium, there has also been substantial progress in the theoretical description of the ultrafast dynamics of atoms and molecules in the gas phase \cite{Remacle,Zewail,FarisG,Gisselbrecht,Henriksen}. At present increasing attention is directed towards the study of surface-adsorbate complexes, since these systems are key to understanding important chemical processes, \cite{Gabor,SurfScience,Hornett,Linic}, and also constitute the next natural paradigm for fundamental light-matter processes. 

However, for surface-adsorbate systems the level of theoretical understanding is at the moment not on the level comparable with free atoms and molecules \cite{Petek1,Frischkorn,Lindroth, Stefanucci1}. Even though there have been recent steps in this direction \cite{Prezhdo,Butriy,Tavernelli,Gross,RvL2,RvLphonon} using sophisticated methods such as Time-Dependent Density Functional Theory (TDDFT) \cite{RG84,EsaRas,Hellgren} and Non-Equilibrium Green's Functions (NEGF) \cite{bookNEG,dipole,Pavlyukh15,Stefanucci1}, the approximate treatment of particle interactions used in practice are not sufficient to accurately describe even the initial stages of desorption. Thus, at present, there is motivation to consider model approaches to ultrafast surface-adsorbate dynamics, that capture the highly non-perturbative situation of correlated electron-nuclear motion and that can be used to gain insight into the outcome of laser induced excitations \cite{KiHyun,Muino}.

Here, we perform an exploratory application of NEGF to adsorbate dynamics, an important physical mechanism
at surfaces out of equilibrium. To this end, we apply NEGF to a simplified version of a recently introduced model 
for adsorbate dynamics induced by ultrashort laser pulses \cite{Bostrom15}. If solved exactly, the model can deal in a simplified way but on equal footing with electron-electron and electron-nuclear interactions, plasmon screening, real-space nuclear dynamics and core-hole relaxation. Thus it permits to study in a coherent way many competing surface response mechanisms and time scales. Here, for simplicity, core levels in the adsorbate and the plasmon response from the substrate are excluded from the outset.

To solve the model exactly, we must limit ourselves to a finite number of substrate sites, or equivalently a limited number of dissipation channels. As an alternative and a way to overcome this obvious limitation, we also consider a semi-infinite substrate within the framework NEGF, where however approximate treatment of interactions must be introduced. For this approach the finite-size version of our model provides an exact benchmark. To disentangle the effects of our various approximations, we perform them in sequence, starting from the numerically exact solution. The first step is to treat the nuclear dynamics in the Ehrenfest approximation while retaining the exact treatment of the electrons. In the second step the electron dynamics is treated within perturbation theory at the level of the Second-Born approximation (2BA). When both approximations are being employed we are free to extend the substrate to become semi-infinite by standard embedding procedures \cite{Petri}. 

It is fair to say that a perturbative treatment of the NEGF (in our case the 2BA) may fail to describe ultrafast real-time dynamics of electrons, especially when dealing with strongly electron correlations. For such systems an accurate and computationally viable first-principles description like NEGF or TDDFT is currently lacking. They both rely on key ingredients [the exchange-correlation (XC) potential for TDDFT and the self-energy $\Sigma$ for the NEGF] that in general are only approximately known and may be not sufficient in the strongly-correlated fast regimes. Recently, we suggested a step towards the solution to this problem using the strengths of both methods, by mixing a non-perturbative adiabatic-TDDFT XC potential with perturbative-based self-energy of NEGF \cite{Hopjan}. Here, we test this hybrid NEGF/TDDFT method at the 2BA level and compare it to the pure NEGF 2BA in our simple model for adsorbate dynamics.

The rest of this paper is structured as follows: the model for adsorbate dynamics is introduced in Sect.~\ref{Model};
the different methods of solutions are described in Sect.~\ref {Exapprox}; the practical details are mentioned in Sect.~\ref {practical};
the results and their discussion are presented in Sect. \ref{numerics}. Some conclusive remarks and outlooks are given in Sect.~\ref{Finish}.

%
%
%
%

\section{Model Hamiltonian for adsorbate dynamics} \label{Model}
The model we use is a simplified version of the one introduced in \tcite{Bostrom15}. 
It consists of a rigid atom chain (the substrate) connected to a mobile adsorbate (Fig.~\ref {fig:system}). The Hamiltonian is
\begin{eqnarray}
\hat{H}(t) = \hat{H}_a + \hat{H}_s + \hat{H}_{as} + \hat{\Lambda}(t), \label{Ham0}
\end{eqnarray}
where $\hat{H}_a$ describes the adsorbate, $\hat{H}_s$ the substrate and $\hat{H}_{as}$ the adsorbate-substrate interaction. To bring the system out of equilibrium a laser field $\hat{\Lambda}(t)$ is applied. 
The adsorbate part $H_a$ is
\begin{align}
H&_{a} = \frac{\hat{p}^2}{2M} + \sum_{v=v_1,v_2}\sum_\sigma \epsilon_v \hat{n}_{v,\sigma} \label{Ham_a} + U \hat{n}_{v_1,\uparrow} \hat{n}_{v_1,\downarrow}
\end{align}
where $\hat{p}$ is the momentum of a mobile atom with mass $M$. The operator $a_{v, \sigma}^\dagger$ creates an electron with spin $\sigma$ at the valence orbital $v$ of the adsorbate with energy $\epsilon_v$, and $\hat{n}_{v,\sigma} = a_{v, \sigma}^\dagger a_{v, \sigma}$. In the adsorbate level $v_1$, valence electrons mutually interact with interaction strength $U$. The substrate is modelled as a non-interacting tight-binding chain with one orbital per site:
\begin{align}
H&_s = - V\!\!\!\sum_{\langle RR'\rangle,\sigma} c_{R,\sigma}^\dagger c_{R',\sigma}\label{Ham_s}~,
\end{align}
where $c_{R,\sigma}^\dagger$ creates an electron at site $R$ with spin $\sigma$. In the following, the hopping amplitude $V$ between neighbouring sites $R$ and $R'$ is taken as the energy unit, i.e. $V = 1$. The adsorbate-substrate interaction Hamiltonian $H_{as}$ is given by 
\begin{align}
H&_{as} = \frac{\kappa}{\hat{x}^4} - g  e^{-\lambda (\hat{x}-1)} \sum_{v,\sigma}\left(a_{v,\sigma}^\dagger c_{S,\sigma} + h.c.\right)  \label{Ham_as}
\end{align}
where we denote the surface site of the substrate by $S$ and $x$ is the adsorbate-surface distance. The first term gives a repulsive interaction between the atomic cores, while the second term is attractive and due to the exchange of electrons between the adsorbate and the surface. We consider $\kappa$ and $\lambda$ as phenomenological parameters that can be adjusted to give reasonable values for the binding energy $E_b$, vibrational frequency $\omega_{ph}$ and effective hopping amplitude $V' = g\langle e^{-\lambda (\hat{x}-1)}\rangle$. 
The shape of the field $\hat{\Lambda}(t)$ in Eq.(\ref{Ham0}) depends on the experiment considered, and is switched on at $t=0$ so that $\hat{\Lambda}=0$ for $t\le0$. In the dipole approximation, and for a field coupling only the adsorbate valence levels, 
$\hat{\Lambda}(t) = \Lambda(t) \sum_{\sigma} a_{v_1\sigma}^\dagger a_{v_2\sigma}+ h.c.$, where 
$\Lambda(t)$ is an arbitrary function of time with a vanishing integral.

%
%
%
%
\section{Exact and approximate treatments} \label{Exapprox}
\subsection{Finite substrate: Exact wavefunction solution}\label{exactsoluz}
An exact solution  of the Schr\"odinger equation for the model presented above is viable only for a short substrate, since the size $\mathcal{S}$ of the Hilbert space $\mathcal{H}$ grows exponentially with the number of orbitals. In more detail, we note that $\mathcal{H} = \mathcal{H}_{el}\otimes\mathcal{H}_{n}$, with $\mathcal{H}_{el}$ the electronic and $\mathcal{H}_{n}$ the atomic (nuclear) Hilbert spaces, respectively. Due the explicit form of $H$ in Eq. (\ref{Ham0}), the size of  $\mathcal{S}_{n}$ grows linearly, while, for a spin-compensated system with $L$ electronic orbitals, we find
$\mathcal{S}_{el}={L \choose L/2}^2 \xrightarrow[ L \to \infty]{} 2^{2L+1}/\pi L$, i.e.
an exponential scaling of $\mathcal{S}$. With and a space-grid
subspace for the nucleus of size $\mathcal{S}_n \simeq 500$, 
substrate lengths $L \lesssim 6$ can be comfortably afforded.

Specifically, the ground state $|g\rangle$ of the model is then determined with exact diagonalization (ED) , by expanding the state vector in the basis $\{ | n_{i\sigma}, x_k\rangle \}$, where $n_{i\sigma}$ is the occupation on site $i$ of electrons with spin $\sigma$ and $x_k$ is the $k$:th mesh-point on a uniform grid in the interval $[0,x_{max}]$. Assuming that the system is in the state $|g\rangle$ at $t=0$, we then propagate the Schr\"odinger equation for $t>0$ using the short iterated Lanczos algorithm.

\subsection{Finite and semi-infinite substrate: NEGF}\label{NEGFsoluz}
If we are not interested in the full wavefunction of the system but we choose to directly look at single particle quantities,
NEGF offer a natural general and in-principle-exact way to go beyond size limitations, and to treat 
electrons and nuclei exactly and on equal footing (in practice approximations must be introduced).
For our present system, this would imply a two-component formulation
of coupled Kadanoff-Baym (KBE) equations, one component for each type of degree of freedom \cite{RvLphonon,Pavlyukh15}. However, to just explore the scope of NEGF for adsorbate dynamics, we here adopt a much 
cruder strategy, namely  we use NEGF for electrons and a classical treatment for the nuclei. The most straightforward approximation is then to use the Born-Oppenheimer separation of electronic and atomic degrees of freedom, and 
for each atomic configuration extract effective potentials generated by the electrons and felt by the atoms, known as Born-Oppenheimer surfaces. Given these potential surfaces, there exists a number of schemes to propagate an initial atomic wave packet; here we consider the Ehrenfest approximation (EA)  \cite{Ehrenfest}, which amounts to replace the nuclear wavepacket with a delta function, and to let the atomic position evolve under the average force from the electrons, or in other words demoting the operators $\hat{x}$ and $\hat{p}$ to classical variables 
\footnote
{Within the Ehrenfest approximation, the size of the Hilbert-space is determined only
by $\mathcal{H}_{el}$, but the substrate-lengths $L$ which can be considered within
our exact treatment remain rather short. Hence the usefulness of NEGF, also
for a classical treatment of the nuclei.}.

\noindent For the model in Eq.~(\ref{Ham0}) the mixed quantum-classical equations in the EA are
\begin{align}
&M \ddot x(t) = -\frac{\partial}{\partial x}\langle\psi_e|\hat{H}(t)|\psi_e\rangle \label{eq:n_ehrenfest}, \\
&i\frac{d}{dt}|\psi_e\rangle = \left(\hat{H}(t)-\frac{p^2}{2M}- \frac{\kappa}{x^4}\right)|\psi_e\rangle \label{eq:e_ehrenfest}.
\end{align}
Once reformulated in the language of NEGF, these become 
\begin{align}
& M \ddot x(t) = \frac{4\kappa}{x^5} - 2g\lambda e^{-\lambda(x-1)}\sum_v \Im\left[G^<_{vS}(t,t)+G^<_{Sv}(t,t)\right], \label{eq:n_negf1} \\
&\left[i\frac{d}{dz_1}-h_{ij}(z_1;x)\right]G_{jk}(z_1,z_2) = \delta_{ik}\delta(z_1,z_2) + \int_\gamma dz \Sigma_{ij}(z_1,z)G_{jk}(z,z_2) \label{eq:e_negf2},
\end{align}
where the single particle Green's function is defined via the time contour ordering $\mathcal{T}$ as $G_{ij}(z_1,z_2) = -i\langle \mathcal{T}c_i(z_1) c_j^\dagger(z_2)\rangle$ for a convenient choice of basis $\{\varphi_i\}$, here taken as the site basis of our model. 
In the coupled equations above, Eqs. (\ref{eq:n_negf1}) governs the nuclear dynamics, whilst Eq. (\ref{eq:e_negf2})
is the standard Kadanoff-Baym equation for $G$, where dependence of the one-body Hamiltonian matrix $h_{ij}$ on $x$ is shown, and an implicit dependence of $\Sigma$ and $G$ on $x$ is understood.

If the system is composed of an interacting region connected to a non-interacting reservoir (as for example, the adsorbate and substrate parts in our system), the latter can be removed from the explicit description and still
taken into account exactly via an embedding procedure \cite{Petri}. In this case, the self-energy for the interacting
region becomes  $\Sigma_{ij}(z_1,z_2) = \Sigma^{em}_{ij}(z_1,z_2) + \Sigma^{MB}_{ij}(z_1,z_2)$. The first term is the embedding self-energy that gives an exact coupling to the reservoir, and the second term is the many-body self-energy that contain all interactions between electrons in the active region. After finding the initial $G_{ij}(t_0,t_0)$ on the imaginary part of the Keldysh contour, the KBE are propagated self-consistently within the same approximation for $\Sigma_{ij}$ \cite{bookNEG}.  With NEGF and the embedding self-energies, a much larger number of orbitals can be explicitly taken into account (plus the levels of the macroscopic reservoir) than e.g. with exact diagonalization, and in any dimensionality. The downside is that interactions now have in general
to be treated in an approximate way, e.g. perturbatively: $\Sigma^{MB}_{ij}(z_1,z_2) \approx \Sigma^{PT}_{ij}(z_1,z_2)$.

\subsection{A NEGF/TDDFT hybrid scheme}\label{NEGFTDDFTsoluz}
In our numerical calculations for adsorbate dynamics, correlations are treated in the 2nd Born approximation (2BA),
$\Sigma^{PT} \equiv \Sigma^{2BA}$. However, starting from this approximation, we also consider a second approach,
recently introduced in \tcite{Hopjan}, which adds to the Kadanoff-Baym equation an exchange-correlation potential $v^{xc,NP}$ of DFT which is nonperturbative in character.  At the
same time, we subtract a perturbative exchange-correlation potential $v^{xc,PT}$, to avoid double counting 
of the interactions.  The approach is conserving in the Kadanoff-Baym sense, and has the same
numerical complexity of standard KBE. In this hybrid NEGF/TDDFT scheme,  the modified Kadanoff-Baym 
equation reads (repeated indices are implicitly summed over)
\begin{align}\nonumber
&\left[i\frac{d}{dz_1}-h_{ij}(z_1;x)
-\left(v^{xc,NP}_{i}(z_1)-v^{xc,PT}_{i}(z_1)\right)\delta(z_1,z_2)\delta_{ij}\right]
G_{jk}(z_1,z_2) \\
&~~~~~~~~~~~~~~~~~~~~~~~= \delta_{ik}\delta(z_1,z_2) + \int_\gamma dz \biggr( \Sigma^{em}_{ij}(z_1,z) + \Sigma^{PT}_{ij}(z_1,z)\biggr)G_{jk}(z,z_2).
\end{align}
Both exchange-correlation potentials are taken in adiabatic local density approximation (ALDA) \cite{ALDA}, 
i.e. the time-dependent XC-potential is approximated as a potential coming from a reference equilibrium system with the same density $v^{xc}_{i}(z_1)\approx{v}{^{xc}_{{ref.}}}(n_i(z_1))$. The reference system is chosen according to the model 
to be solved, as discussed in the next section.%

\subsection{Additional aspects of the NEGF/TDDFT scheme}
\begin{figure}[t]
        \hspace{-0.8cm}
    \begin{minipage}{.5\textwidth}
        \centering
        \includegraphics[width=100mm]{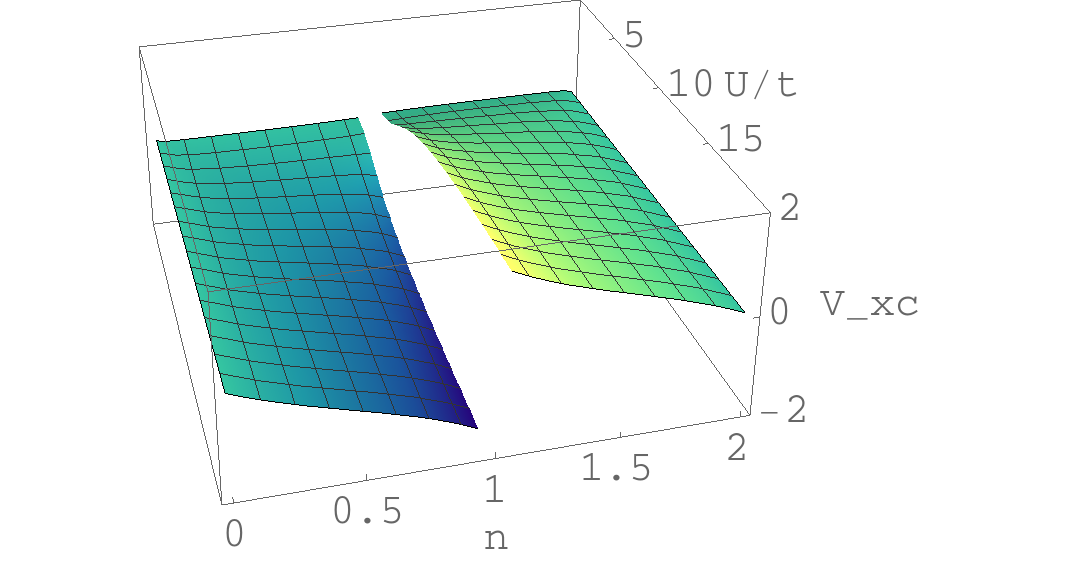}
    \end{minipage}%
    \hspace{0.4cm}
    \begin{minipage}{.5\textwidth}
        \centering
        \includegraphics[width=100mm]{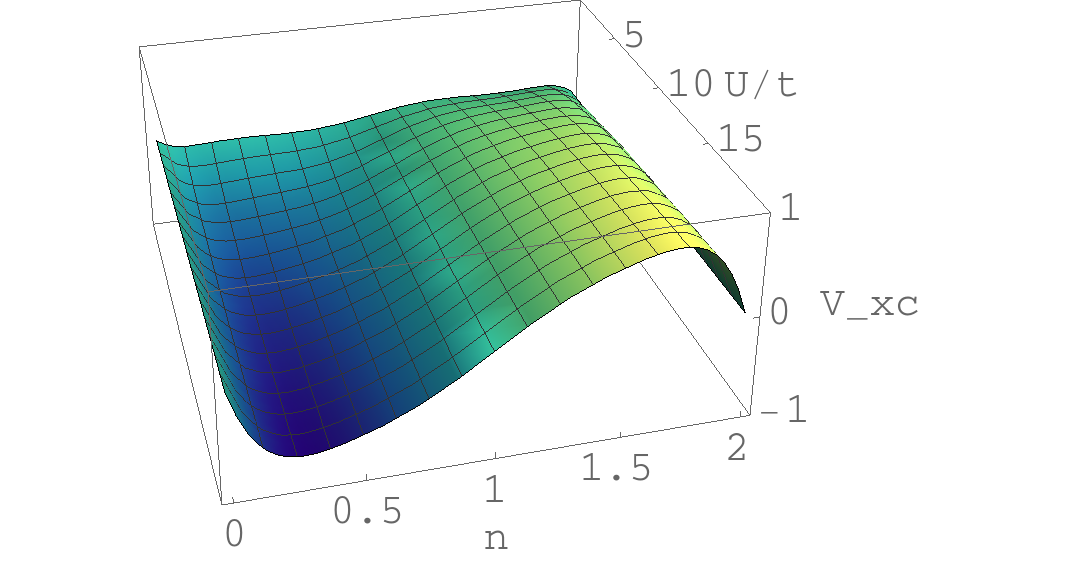}
        \label{fig:BALDA}
    \end{minipage}%
\caption{ XC potentials for the reference homogeneous one-dimensional Hubbard model $H=t\sum_{\langle ij\rangle,\sigma }a^{\dagger}_{i,\sigma}a_{j,\sigma}+U\sum_{i}n_{i,\downarrow}n_{i,\uparrow}$. The exact nonperturbative exchange-correlation potential $v_{ref.}^{xc,NP}$ is shown to the left, and to the right is the corresponding perturbative potential $v_{ref.}^{xc,PT}$ in the 2BA. We plot the potentials as a function of uniform density $n$ and the ratio of interaction and hopping $\frac{U}{t}$ for fixed interaction $U=4$.}
\label{fig:potentials}
\end{figure}

In the NEGF/TDDFT method, we need suitable XC potentials 
$v^{xc,NP}$ and $v^{xc,PT}$ for the lower valence level. The geometrical structure of the model in Fig.~\ref{fig:system} does not allow for a direct application of standard 1D ALDA (which is obtained from a 1D homogeneous Hubbard model
where each site has a bond on the left and one on the right). However, both in equilibrium and for time-varying perturbations acting only on the adsorbate, symmetry arguments permit to re-interpret the adsorbate as a 
two-orbital central site connected to the parity-even states (but not to odd ones) of two identical 1D semi-infinite substrates, with a reduced site-substrate(s) coupling $V'_{\rm sym}=V'/\sqrt{2}$.
For the lower valence level $v_1$ of the central site in the symmetric model, we employ the ALDA potential $v^{xc}\approx{v}{^{xc}_{{ref.}}}\left(n_{\nu_1},U/V'_{\rm sym}\right)$ using the homogeneous 1D Hubbard model as the reference equilibrium system \cite{GunSchon,Capelle,CV2008}. The corresponding potentials $v^{xc,NP}$ and $v^{xc,PT}$ are shown in Fig. \ref{fig:potentials}: they mutually differ in many ways in their behavior, the key difference being that $v^{xc,NP}$ ($v^{xc,PT}$) has (has not) a discontinuity at half-filling.
%
%
%
\section{Practical details}\label{practical}
\subsection{On the different treatments} 
In the actual calculations, we consider both finite ($L=5$)  and  semi-infinite 1D substrates (see Eq.~\ref{Ham0} and Fig.~\ref{fig:system}). 
For $L=5$, the model is solved exactly with a full-quantum mechanical solution for the electrons and the nucleus (Sect.~\ref{exactsoluz}) and, in the EA, with different degrees of approximations for the electron dynamics (Sects.~\ref{NEGFsoluz},~\ref{NEGFTDDFTsoluz}).
In the EA for the nuclei plus the exact dynamics of the electrons (EA+Ex$_e$), the electronic ground state $|g_e\rangle$ and nuclear equilibrium position $x_{eq}$ are found at the stationary points of Eq.~(\ref{eq:n_ehrenfest}-\ref{eq:e_ehrenfest}), after which we propagate Eq.~(\ref{eq:n_ehrenfest}-\ref{eq:e_ehrenfest}) in time using the Lanczos technique for the electrons and a velocity Verlet-type algorithm for the nuclei. When the EA is used in conjunction with NEGF, the electronic correlations are taken into account at the 2BA level (EA+2BA), or via the hybrid NEGF/TDDFT scheme (EA+Hyb). The ground state Green's function $G_{ij}(t_{0},t_{0})$ and equilibrium position $x_{eq}$ are now found solving self-consistently Eq.~(\ref{eq:e_negf2}) on the vertical track of the contour for $G$ and Eq.~(\ref{eq:n_negf1}) for $x$.
For the dynamics, Eqs.~(\ref{eq:n_negf1}-\ref{eq:e_negf2}) are propagated in time according to standard procedures on the time-square for $G$ \cite{Kohler,Stan,Puig} together with a velocity Verlet algorithm for the nucleus.  Within the NEGF treatments, we also consider a semi-infinite 1D substrate  via an embedding self energy \cite{Petri}, with the solution method otherwise unchanged. 

\subsection{On the external fields} 
We used two time-dependent fields to perturb the system:
$\Lambda(t)\equiv \Lambda_a(t)=Ae^{-(t-t_0)^2/2\sigma^2} $ and 
$\Lambda(t)\equiv \Lambda_b(t)= \Lambda_{a}(t)\sin\left(\omega t +\pi/2\right)$,
with
amplitude either $A=2$ or $4$. The electronic parameters of the model are chosen as $V=1$, $\epsilon_1 = 1$, $\epsilon_2 = 3$, thus giving a cationic adsorbate in the ground state, and we consider the interaction strengths $U=4,6$ and $8$. For the nucleus, $\kappa = 2.4$, $g = 1.8$ and $\lambda = 1.2$, 
giving a binding energy $E_b \simeq 1.5$, an harmonic frequency $\omega_{ph} \simeq 0.2$ and an effective hopping amplitude $V' \simeq 1.8$. These values are typical for chemisorption.

The Gaussian pulse $\Lambda_{a}$ is an artificial perturbation and serves as a tool to investigate slower dynamics, while the sinusoidally modulated pulse $\Lambda_{b}$ represents a realistic pulse with a carrier wavelengh of $400$nm (for a hopping amplitude $V \simeq 2$ eV the unit of time is $1/V \simeq 1.3$ fs, which is the period of $400$ nm radiation). Both perturbations extend from $t=0$ to $t=8$, with time measured in units of the inverse hopping $1/V$. In this time interval the evolution is dominated by the perturbation and all the methods give
basically the same result. For later times the dynamics is dominated by the memory of the system, and the solutions differ depending on the approximation used.
\begin{figure}[h]
    \centering
    \includegraphics[scale=1]{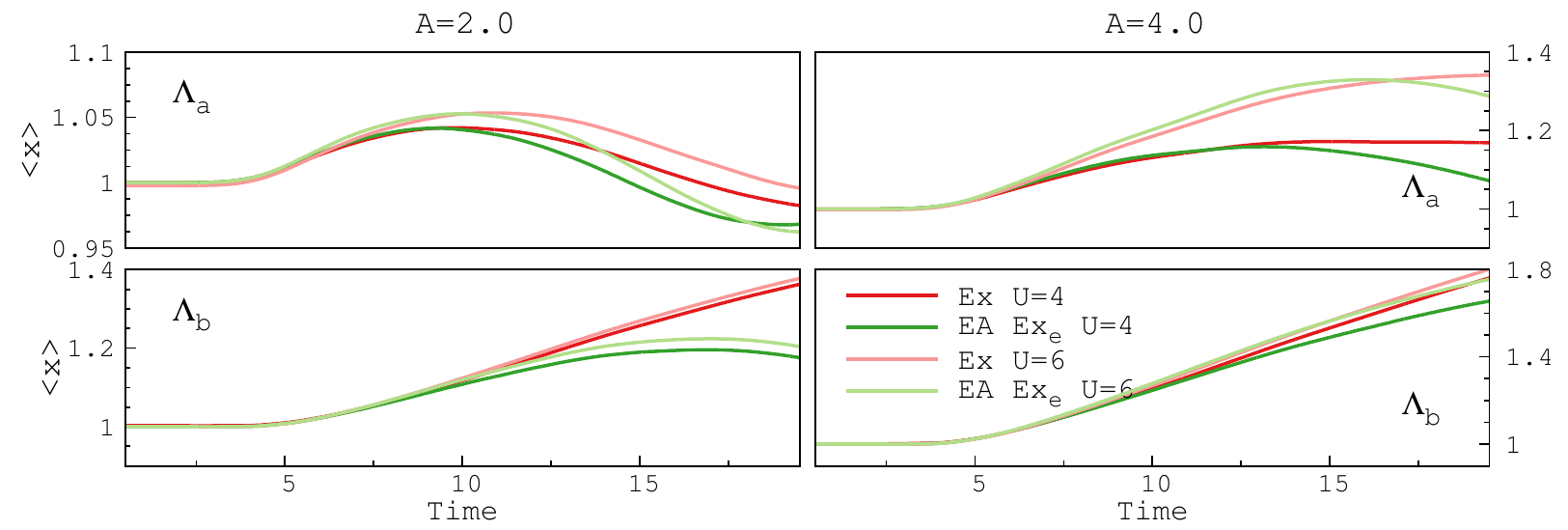}
\caption{Time-dependent interatomic distances in the exact quantum treatment (red) and the EA+Ex$_e$ (green), for the values $U=4$ and $U=6$ of the interaction strength. The first row shows results for the external field $\Lambda_a(t)$ of amplitude $A = 2.0$ and $A = 4.0$, and the second row gives the corresponding results for the field $\Lambda_b(t)$.}
\label{fig:position}
\end{figure}

\section{Results}\label{numerics}
\subsection{Exact vs Ehrenfest dynamics for the nuclei} 

Let us first look at the evolution of the nuclear subsystem, as shown in Fig.~\ref{fig:position}. With the full quantum mechanical solution we observe bond stretching for the pulse $\Lambda_a$, but for large times the nuclear position returns to a region close to its equilibrium value. For the pulse $\Lambda_b$ the nuclear coordinate continues to increase also for large times, signalling a finite probability of desorption. This is typical of the full quantum solution, where in general one part of the initial wave packet remains bound and performs oscillations in the potential well, while another part overcomes the potential barrier and thereafter propagates quasi-freely (strictly the potential is zero only for $x\to \infty$). We note that with increasing interaction $U$ and increasing amplitude $A$, the desorption probability also increases. The former is a result of the interaction induced quenching of electronic motion between the adsorbate and surface, resulting in a lower average bond kinetic energy and therefore a weaker bond (cf. the form of electron-nuclear coupling in Eq.~\ref{Ham_as} and the definition $K_{as} = c_{v_1}^\dagger c_S + c_{v_2}^\dagger c_S + h.c.$ for the bond kinetic energy).

Since the EA is a classical treatment of nuclei the only possible outcomes of an experiment is to have either full or no desorption, and in the intermediate regime the approximation will inevitably fail to reproduce the exact dynamics. This is verified in Fig.~\ref{fig:position} where the exact results are best reproduced in the cases where the quantum solution predicts either a very small or very large desorption probability. The agreement is in general better for the pulse $\Lambda_a$, which varies more slowly in time and is closer to the adiabatic situations where the EA is known to perform well. 

\subsection{Exact vs NEGF electronic correlations within the Ehrenfest dynamics.} 

\begin{figure}[h]
    \centering
    \includegraphics[scale=1]{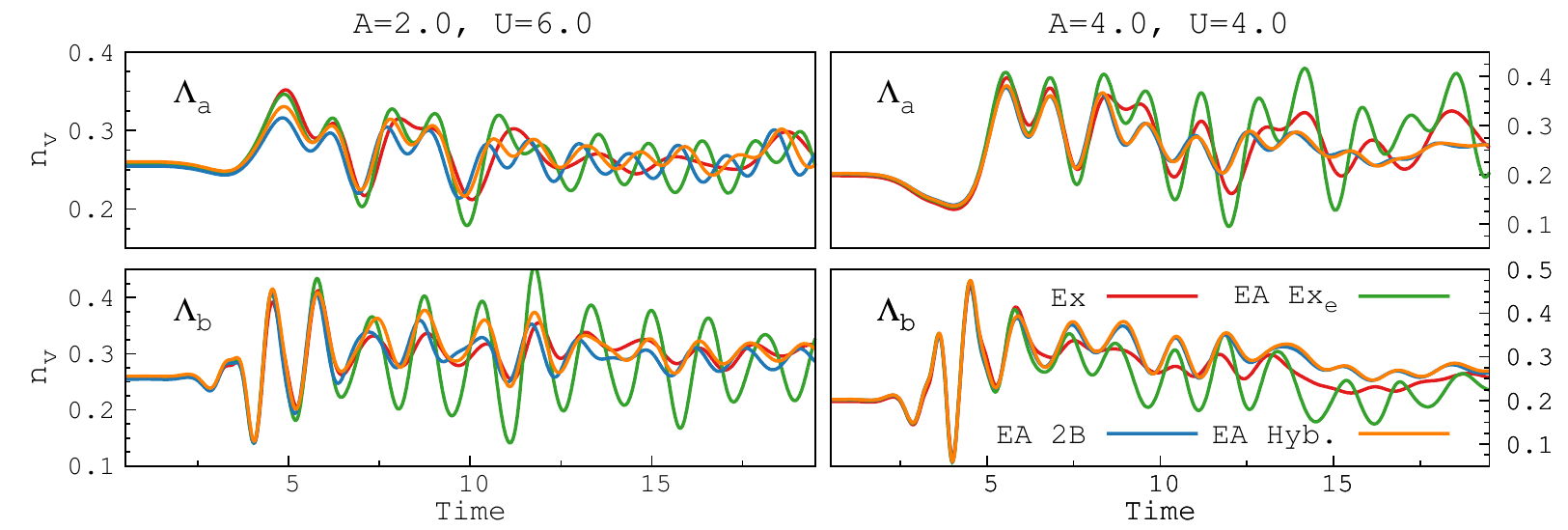}
\caption{Time-dependent densities of the lower valence level in the exact quantum treatment (red) and the approximate schemes EA+Ex$_e$ (green), EA+2B (blue) and EA+Hyb (orange).The first row shows results for the external field $\Lambda_a(t)$ of amplitude $A = 2.0$ with interaction strength $U= 6$ and $A = 4.0$ with interaction strength $U=4$, and the second row gives the corresponding results for the field $\Lambda_b(t)$.}
\label{fig:finiteelectron}
\end{figure}

We now turn our attention to the electron dynamics, as depicted in Fig.~\ref{fig:finiteelectron}. When comparing the exact solution with the EA+Ex$_e$ dynamics we see clear differences, that are most pronounced in the case of partial desorption. In particular the EA shows oscillations in the electron density of greater amplitude, that persist even for long time. This can be explained in part by the inability of the EA to transfer energy from the electronic to the nuclear subsystem, due to neglect of electron-nuclear correlations, that in the exact simulation act as a dissipation channel for the electrons. In part it is also due to the fact that the effective adsorbate-surface hopping amplitude is larger in the EA, making is more probable for the electrons to tunnel between the substrate and the adsorbate. With quantum mechanical nuclei, the hopping amplitude (proportional to $e^{-\lambda x}$, see Eq.~\ref{Ham_as}) is decreased by the presence of a split-off part of the nuclear wave packet having propagated to large distances. 

Comparing the exact and the EA+2B electron evolution we observe that the oscillations in density seen in the EA are reduced compared to the EA+Ex$_e$ result, and that this difference increases with stronger interactions. The damping of electron oscillations is a generic feature of approximate solutions of the KBE in finite systems, and was first discussed in \cite{Puig}. Briefly, the reason for damping is that self-consistent approximations (like the 2B) include infinitely many electron-hole excitations (through the infinite set of diagrams contained in the self-energy), which cannot occur in finite systems. Thus, even though the agreement between the exact and the EA+2B solution seems better than EA+Ex$_e$, it should be kept in mind that this result is due to unphysical effects. 

With the EA+Hyb method we observe a slightly better agreement in phase with the EA+Ex$_e$ treatment, but for the model considered here the improvement of local correlations brought by the exchange-correlation potential have a rather small overall effect. This points to the importance of non-local correlations for the desorption process, and to capture more closely the exact dynamics approximations beyond 2B (or even inclusion of non-perturbative effects) are necessary.  As a first step in this direction we are currently working on including correlations in the $T$-matrix approximation.

\subsection{The effect of a semi-infinite substrate} 
\begin{figure}[h]
    \centering
    \includegraphics[scale=1]{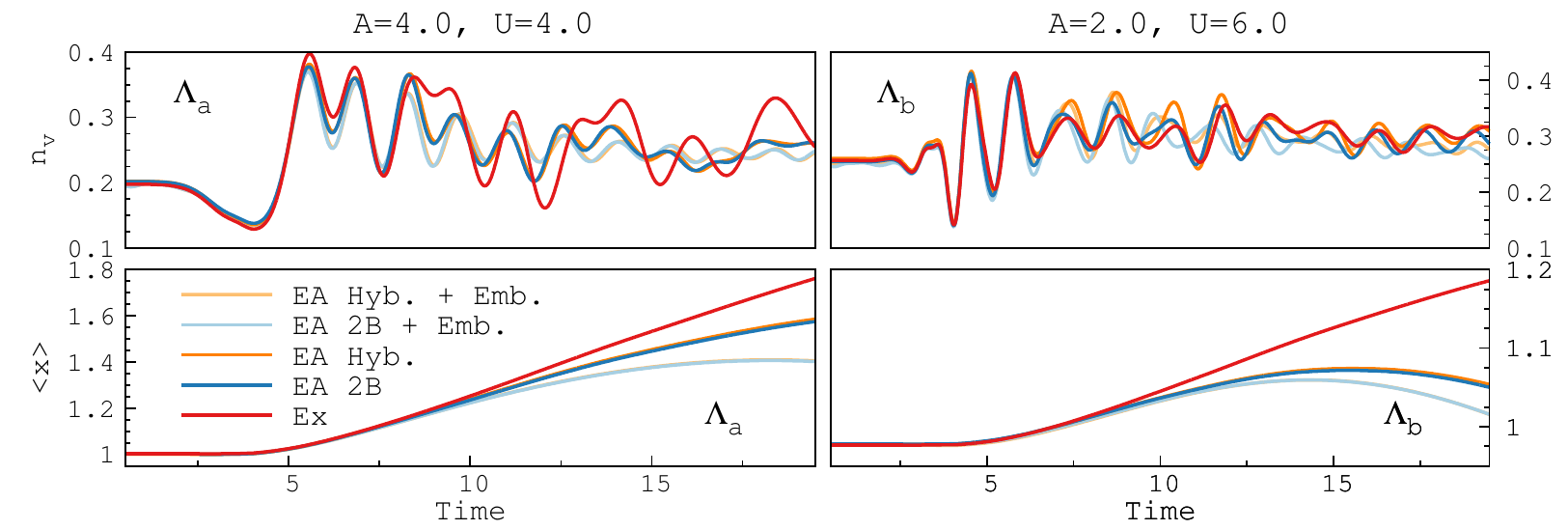}
\caption{Time-dependent densities of the lower valence level (top) and interatomic distance (bottom), for a finite chain in the exact quantum treatment (red) and the approximate schemes EA+2B (blue) and EA+Hyb (orange), as well as for a contacted system within EA+2B (light blue) and EA+Hyb (light orange). The first column shows results for the external field $\Lambda_a(t)$ of amplitude $A = 4.0$ with the interaction strength $U=4$, and the second column gives the results for the external field $\Lambda_b(t)$ of amplitude $A = 2.0$ with the interaction strength $U=6$.}
\label{fig:extendedelectron}
\end{figure}

Finally, we contact the system to a semi-infinite lead and compare the resulting dynamics within EA+2B and EA+Hyb to the corresponding finite system simulations (see Fig.~\ref{fig:extendedelectron}). Interestingly, the addition of the particle reservoir seems to have a rather small impact on the short-time dynamics, both for the electronic and nuclear subsystem. At larger times the desorption probability decreases in the presence of a lead, since there is now an additional energy dissipation channel for the electrons that reduce the amount of energy transferred to the nuclei. However, for the nuclear part there is not much difference between the EA+2B and EA+Hyb methods.

The electron dynamics of the finite and extended systems show a qualitative agreement, the major difference being that the amplitude of the density oscillations decay faster in the latter case. For a contacted system the damping of electron oscillations is an inherent physical effect, in contrast to the artificial behavior discussed in the previous section. For the faster pulse $\Lambda_b$ we also observe a slight dephasing of the EA+2B and EA+Hyb solutions, favoring the EA+Hyb method in both the finite and contacted case.

The difference between the EA+2B (or EA+Hyb) and the exact solution is typically greater than the difference between the finite and contacted solution within a given approximation scheme. This might imply that to model adsorbate-surface systems, and in particular non-perturbative processes such as desorption, it is more important to properly account for the electron-electron and electron-nuclear correlations close to the surface, than for the infinite extent of the system. This conclusion is strongly dependent on the parameters chosen, but for a strong electron-nuclear coupling (so that the system is in the surface molecule limit) we expect it to hold.

%
%
%
%

\section{Conclusions}\label{Finish}
We present a model to treat real-time surface-adsorbate system dynamics induced by ultrafast laser pulses, and solve it exactly numerically in its finite size realisation. The exact results are compared to a treatment within the framework of NEGF, on the level of the second Born approximation and also to a recently introduced hybrid NEGF/TDDFT approach. By the standard embedding procedure this allows us to also solve the semi-infinite version of our model. In the finite size case we treat the nuclear coordinate exactly using first quantisation, and compare it to the semi-classical case of having exact electronic dynamics and nuclear motion within the Ehrenfest approximation.

For the process of desorption, interesting due to its non-perturbative character, we find good qualitative agreement between the exact and Ehrenfest nuclear dynamics when the exact result predicts either a very small or a very large desorption probability. In the intermediate case, characterized by a splitting of the nuclear wave packet into a bound and a quasi-free part of comparable magnitude, the Ehrenfest treatment fails to reproduce the exact results. For the case of chemisorption considered here, we further find that the inclusion of a semi-infinite substrate have a limited effect but predicts a slightly lower desorption probability. Our results also indicate the importance of non-local correlation effects beyond second Born.

%
%
\section*{References}
\bibliographystyle{iopart-num}

\end{document}